\documentclass[aps,pra,superscriptaddress,twocolumn,showpacs]{revtex4-1}

\usepackage{amsmath}    
\usepackage{amsfonts}
\usepackage{bm}
\usepackage{amssymb}
\usepackage{appendix}
\usepackage{graphicx}   
\usepackage{subfigure}  
\usepackage{color}

\newcommand{\re}{\textrm{Re}}
\newcommand{\im}{\textrm{Im}}

\newcommand{\gper}{\gamma_\perp}
\newcommand{\gpar}{\gamma_\parallel}
\newcommand{\gse}{\gamma_{\textrm{SE}}}

\begin{document}

\title{Steady-state ab initio laser theory for complex gain media }
\author{Alexander Cerjan} 
\affiliation{Department of Applied Physics, Yale University, New Haven, CT
06520, USA}
\author{Y.~D.~Chong}
\affiliation{School of Physical and Mathematical Sciences, Nanyang Technological University, Singapore 637371, Singapore}
\author{A.~Douglas Stone}
\email[]{douglas.stone@yale.edu}
\affiliation{Department of Applied Physics, Yale University, New Haven, CT
06520, USA}

\date{\today}

\begin{abstract}
We derive and test a generalization of Steady-State Ab Initio
Laser Theory (SALT) to treat complex gain media.  The generalized theory (C-SALT)
is able to treat atomic and molecular gain media with diffusion and multiple 
lasing transitions, and semiconductor gain media in the free carrier approximation
including fully the effect of Pauli blocking.  The key assumption of the theory is 
stationarity of the level populations, which leads to coupled self-consistent equations
for the populations and the lasing modes that fully include the effects of openness and non-linear spatial hole-burning.
These equations can be solved efficiently for the steady-state lasing properties by a similar
iteration procedure as in SALT, where a static gain medium with a single transition is assumed.
The theory is tested by comparison 
to much less efficient Finite Difference Time Domain (FDTD) methods and excellent agreement is found.  
Using C-SALT to analyze the effects of varying gain diffusion constant we demonstrate 
a cross-over between the regime of strong spatial hole burning with multimode lasing to a regime of
negligible spatial hole burning, leading to gain-clamping, and single mode lasing.  The effect of
spatially inhomogeneous pumping combined with diffusion is also studied and a relevant length
scale for spatial inhomogeneity to persist under these conditions is determined.  For the semiconductor gain model,
we demonstrate the frequency shift due to Pauli blocking as the pumping strength changes.

\end{abstract}


\maketitle

\section{Introduction}

Laser theories describe the properties of self-organized electromagnetic oscillators based on stimulated emission in an
inverted gain medium.  In semiclassical laser theory the electromagnetic equations are
simply Maxwell's equations, typically in the form of a Helmholtz equation, coupled to matter through 
a polarization or susceptibility which describes the pumped, inverted gain medium.  The susceptibility is
sometimes modeled as a classical charged oscillator, but more typically is calculated quantum mechanically.
The use of Maxwell's equations implies that quantum fluctuations of the electromagnetic fields are neglected
at this level of the theory.  In the full quantum theory of the laser the fields are represented by operators and
calculation of their higher moments yields the fluctuation properties.  However most of the basic properties of 
interest for a laser are determined by the semiclassical laser equations, which in addition to Maxwell's equations
include equations for the polarization and atomic populations of the gain medium (the diagonal and off-diagonal
matrix elements of the density matrix).  These properties include
the lasing thresholds, modal intensities, modal frequencies, and spatial variation of the internal fields and output intensities, 
as well  as the spatial variation of the inversion of the gain medium and some other dynamical properties such as the
relaxation oscillation frequency and decay rate of each lasing mode.  The full semiclassical laser equations are coupled
non-linear partial differential equations in space and time and are not easily solved even numerically.
In typical textbook treatments the spatial variation of the quantities in these equations is either completely neglected or
treated through an approximate expansion in closed cavity states which are taken to be orthogonal, 
hence turning the resulting equations into coupled ordinary non-linear differential equations in time. Furthermore,
the gain medium is often described as a simple two-level atomic system and modeled using the Bloch equations.

The past two decades have seen an explosion in the types and configurations
of novel laser systems. This development has been fueled by advances in microfabrication
techniques and motivated by applications to integrated on-chip optics,
for more efficient optical communications and sensing, as well as basic scientific interest.
Many of these advances involve lasers with complex gain media, such as
semiconductor lasers \cite{koch89,koch}, quantum cascade lasers \cite{faist94,gordon08}, 
and rotationally excited gases \cite{chua11}, for which the two level atomic
system approximation is poorly suited. For example, though the band structure
of the semiconductor gain medium can be approximated as a series of two
level atomic transitions, multiple transitions are required to represent
the effects of Pauli blocking \cite{ho06}. Cascaded-transition quantum cascade lasers
are designed with two lasing transitions to operate at longer wavelengths \cite{capasso98,huang13}.
Additionally, for both rotationally
excited gases and semiconductor media the carriers are allowed to diffuse
through the cavity, an effect that is not addressed in the Maxwell-Bloch equations.
Traditional methods for simulating lasers, such as Finite-Difference
Time-Domain (FDTD) calculations, are too computationally demanding to study
a large parameter space of the system, particularly if it is necessary to include
carrier diffusion effects \cite{jungo04,pernice07,hess08},

Recently, a theoretical approach has been proposed by T\"{u}reci, Stone, \textit{et al.}\ to
obtain directly the steady state solutions for complex laser systems without time integration \cite{tureci06,tureci07,tureci08,ge10}.
This theory, the Steady-state Ab initio Laser Theory (SALT), employs only a single 
important approximation, the stationary inversion approximation (SIA), to treat multimode lasing, and
no significant approximation in treating single-mode lasing, and it treats the spatial degrees of freedom
exactly.  The SALT equations are frequency domain wave equations for the lasing modes, coupled non-linearly
through the spatially varying cross gain saturation.  These equations can be solved
efficiently for the steady-state properties of laser cavities of arbitrary complexity, 
in any number of dimensions, using
a specialized non-hermitian basis set which will be introduced below.
This approach allows a clearer and partially analytic understanding of the lasing solutions, and 
has already led to a number of new discoveries, such as mode frustration in
partially pumped cavities \cite{liertzer12}, control of emission properties of random lasers through
selective pumping \cite{andreasen_creation_2009,andreasen_effects_2010,ge_unconventional_2011,bachelard_taming_2012,hisch_pump-controlled_2013}, 
and a more general form of the Schawlow-Townes
linewidth formula \cite{chong12,pillay_2014}.  For single transition gain media without diffusion, 
SALT has also been shown to give excellent agreement
with FDTD simulations at a substantially reduced computational cost \cite{ge08,cerjan12}.
In the case of a single transition the gain susceptibility for N-level lasers can be explicitly calculated in terms of
the stationary inversion profile and the lasing fields and inserted into the wave equation so as to require only a single set of coupled electric field equations \cite{cerjan12}.  It can then be shown that the N-level model
can be reduced to the two-level Maxwell-Bloch model with renormalized relaxation and pump parameters
\cite{cerjan12}.

Such a simplification is not possible with multiple lasing transitions, but, as we will show,
a stationarity assumption (The Stationary Population Approximation) which is only slightly
more general than that used in SALT, leads instead to coupled sets of
field and population equations that can be solved iteratively almost as 
efficiently as the SALT equations. The generalized theory, which we call complex-SALT
(C-SALT), allows us to treat steady-state lasing with (i) an arbitrary number of atomic levels and lasing
transitions, and (ii) gain diffusion. As already noted, direct integration in space and time of
the lasing equations with diffusing gain centers \cite{jungo04,pernice07,hess08} is very challenging.
C-SALT can obtain the same steady-state solution with relative ease, allowing one to use
the method for exploration of basic laser physics, and, potentially, for device design.

Below we first solve the C-SALT equations for multimode lasing with multiple transitions but
without gain diffusion and demonstrate excellent
agreement with FDTD simulations. Including gain diffusion in the C-SALT equations
and studying both 1D and 2D lasers cavities, we identify
two distinct physical situations in which carrier diffusion can have a substantial effect.
The first is when the carrier diffusion competes with spatial
hole-burning in a uniformly pumped cavity, leading to a transition between multimode lasing
to a gain-clamping regime in which only a single lasing mode reaches threshold (as one 
finds when spatial hole-burning is absent \cite{haken}).  
The second situation is when carrier diffusion competes with non-uniform pumping
so as to mitigate the spatial selection effects which occur in the absence of diffusion.
We are not aware of prior quantitative theoretical studies of these effects.

The remainder of this paper is organized as follows. In section II we introduce
the semiclassical lasing equations and analyze the most general conditions under
which they lead to steady-state multimode lasing.   
Section III derives the equations for the steady-state (A) for a single transition, leading
to the SALT equations (B),(C) for multiple lasing transitions and gain diffusion, leading
to the C-SALT equations, which are the main result of this work.
Section IV demonstrates quantitative
agreement between SALT and FDTD simulations as well as showing the
transition between spatial hole-burning and gain clamping as the
diffusion constant is increased.  In section V a discussion
of the Stationary Population Approximation is presented. Section VI
demonstrates how SALT can be used to simulate lasers described by a
free-carrier semiconductor model. Finally, a conclusion is given in
section VII.

\section{Overview of the semiclassical lasing equations}

As noted above, the semiclassical lasing equations neglect the operator nature of the electromagnetic
fields, so that the lasing fields are determined by Maxwell's wave equation (coupled to a quantum gain
medium),
\begin{equation}
\left[ \nabla \times \nabla \times - \frac{\varepsilon_c}{c^2}(\mathbf{x}) \partial_t^2 \right]\mathbf{E}(\mathbf{x},t) = \frac{4\pi}{c^2} \partial_t^2 \mathbf{P}_g(\mathbf{x},t), \label{maxEqn}
\end{equation}
where the effect of the passive cavity is described by the linear
cavity dielectric function, $\varepsilon_c(\mathbf{x})$, which in general varies in space and frequency (although
typically we will neglect the frequency variation).  The amplifying response of the gain medium is described by the
non-linear gain polarization, $\mathbf{P}_g$, which acts as a source for this equation and includes contributions
from all of the lasing transitions of every gain atom in the cavity. At this stage we will assume that the gain medium
arises from independent identical atomic, molecular or defect centers; the case of semiconductor gain media with band 
excitations acting as gain centers will described in Section  \ref{semiSec} below.

The gain polarization can be expressed as
\begin{equation}
\mathbf{P}_g(\mathbf{x},t) = -\sum_{\alpha} \delta(\mathbf{x} - \mathbf{x}^{(\alpha)}) \textrm{Tr}[\hat{\rho}^{(\alpha)} e \hat{\mathbf{x}}_\alpha],
\end{equation}
in which the index $\alpha$ runs over all of the gain atoms in the cavity at locations
$\mathbf{x}^{(\alpha)}$, $e$ is the charge of an electron, 
and the $M \times M$ density matrix of atom $\alpha$ is denoted by
$\hat{\rho}^{(\alpha)}$, where $M$ is the number of atomic levels involved in the lasing process.
Among the $M$ levels a subset of them will contribute to lasing and support lasing transitions; the others
will simply be part of the downward cascade of electronic excitations involved in the pumping and emission
in steady-state.  Lasing transitions will arise from pairs of level which have sufficiently large polarization
due to their population inversion to contribute substantially to a lasing line at a nearby frequency.
These pairs have a dipole moment,
\begin{equation}
\boldsymbol{\theta}_{nm}^{(\alpha)} = e \langle n | \hat{\mathbf{x}}^{(\alpha)} | m \rangle.
\end{equation}
The dipole moment is always zero for $n=m$ due to spatial symmetry.
By assuming that the full polarization can be expressed in term of the density matrix for individual atoms, we are 
ignoring interatomic coherence effects which are quite small for conventional lasers (but not e.g.\ for polariton lasers).

To complete the semiclassical lasing equations, one must consider the quantum
equations of motion for the average polarization of the gain medium,
\begin{equation}
\partial_t \mathbf{P}_g (\mathbf{x},t) = -N(\mathbf{x}) \sum_{n}^M \sum_{m}^M \partial_t(\rho_{nm}) \boldsymbol{\theta}_{mn},
\end{equation}
where we have now assumed identical ``atoms" and written the density matrix elements as 
$\rho_{nm} = \langle n | \hat \rho | m \rangle$,
and we initially assume a fixed density of gain atoms $N(\mathbf{x})$.
The evolution of the density matrix can be found from the Heisenberg equation of motion,
\begin{equation}
\partial_t \rho_{nm} = \frac{-i}{\hbar} \langle n | \left[ H_0 + H_I, \hat{\rho} \right] |m \rangle,
\end{equation}
where the atomic Hamiltonian $H_0 | n \rangle = E_n | n \rangle$, and the interaction Hamiltonian
can be written as $H_I = e \hat{\mathbf{x}} \cdot \mathbf{E}(\mathbf{x},t)$. Upon evaluating the commutator 
and simplifying, the evolution of the density matrix elements can be re-written as
\begin{equation}
\partial_t \rho_{nm} = -i\omega_{nm} \rho_{nm} - \frac{i}{\hbar} \sum_k^M \left(\boldsymbol{\theta}_{nk} \rho_{km} - \rho_{nk} \boldsymbol{\theta}_{km} \right) \cdot \mathbf{E}(\mathbf{x}, t), \label{denMatEq}
\end{equation}
where $\omega_{nm} = (1/\hbar)(E_n - E_m)$ is the transition frequency.

From this equation, we can see that off-diagonal density matrix elements, which determine the
gain polarization, can couple to one another within manifolds of atomic transitions.  If we include
all terms in the equation of motion, then the time evolution of a specific off-diagonal element, $\rho_{nm}$,
will depend not only on the level populations $(\rho_{nn}, \rho_{mm})$, but also on other off-diagonal elements, i.e.\
on the polarization of other transitions.  If this is the case one cannot arrive at lasing equations of
the standard form and one cannot define the polarization associated with a specific transition. 
However, physically these off-diagonal terms correspond to coherent multiple excitation, leading 
to effects such as electrically induced transparency, and inversion-less lasing. Conventional lasers do
not typically operate in this regime. To reach this regime, the non-radiative relaxation rates between
non-lasing transitions must be of similar order to those between lasing levels, which makes
it difficult for the gain medium to build up the necessary inversion to lase \cite{haken87}.
As such, we will assume that we are in the weakly coupled polarization regime and
thus that off-diagonal density matrix elements depend only on the level populations of that specific 
pair of levels. The equation of motion for the off-diagonal elements for such a pair is
\begin{align}
\partial_t \rho_{nm}  = & -\left(\gamma_{\perp,nm} + i\omega_{nm}\right) \rho_{nm} \notag \\
&+ \frac{i}{\hbar} (\rho_{nn} - \rho_{mm}) \boldsymbol{\theta}_{nm} \cdot \mathbf{E}(\mathbf{x}, t), \;\;\;(m\ne n) \label{offDiagEqn}
\end{align}
where we have now added the effect of environmental dephasing on the gain atoms in the standard manner in terms of a transverse relaxation/dephasing rate $\gamma_{\perp,nm}$.
 
With the above assumption, the total gain polarization can now be broken up into $N_T$ constituent
polarizations of each lasing transition,
\begin{align}
\mathbf{P}_g(\mathbf{x},t) =& \sum_j^{N_T} \mathbf{p}_j(\mathbf{x},t) \label{sepPolEqn} \\
\mathbf{p}_j^+(\mathbf{x},t) =& -N(\mathbf{x}) \rho_{nm} \theta_{mn}
\end{align}
where each transition, previously labeled by the pair of levels, $n,m$, 
has been relabeled with a transition index $j$, where $\mathbf{p}_j^+$ is the positive frequency
part of $\mathbf{p}_j = \mathbf{p}_j^+ + \mathbf{p}_j^-$, and by definition $\omega_{nm} > 0$. 

From the form of the gain polarization as a transition sum, Eq.\ (\ref{sepPolEqn}),
it is clear that every constituent polarization in the gain medium contributes
to the total source term in the wave equation, (\ref{maxEqn}). Strictly speaking, it
is thus impossible to say that one portion of the electric field is driven by only a
single transition if multiple transitions are present in the gain medium, though practically
there are many cases where these transitions are so well separated in frequency that this is effectively
the case. Each constituent polarization, from Eq.\ (\ref{offDiagEqn}), obeys its own equation of motion,
\begin{equation}
\partial_t \mathbf{p}_j^+(\mathbf{x},t) = -\left(\gamma_{\perp,j} + i \omega_{a,j}\right) \mathbf{p}_j^+ - \frac{i d_j}{\hbar}\left(\boldsymbol{\theta}_j \cdot \mathbf{E} \right) \boldsymbol{\theta}_j^*, \label{sepPolSolved}
\end{equation}
in which the properties of the $j$th constituent polarization are given in terms of the
dephasing rate, $\gamma_{\perp,j}$, the atomic transition frequency, $\omega_{a,j}$,
the constituent inversion, $d_j = N(\mathbf{x})(\rho_{nn}^{(j)} - \rho_{mm}^{(j)})$, and the dipole matrix element, $\boldsymbol{\theta}_j$.

The equation of motion for the density matrix, Eq.\ (\ref{denMatEq}), also determines
the evolution of the populations in each atomic level, given by the diagonal
elements of the density matrix,
\begin{equation}
\partial_t \rho_{nn} = -\frac{i}{\hbar} \sum_k^M \left(\boldsymbol{\theta}_{nk} \rho_{kn} - \rho_{nk} \boldsymbol{\theta}_{kn} \right) \cdot \mathbf{E}(\mathbf{x}, t).
\end{equation}
As can be seen here, atomic level populations couple only to the constituent polarizations with
which they share a transition, and atomic levels which are not a part of any lasing
transition do not appear at this stage on the right hand side of the equation for the populations.
However typically all levels are coupled non-radiatively through other degrees of freedom
(``the bath") and these effects need to be included phenomenologically in the standard manner, leading to
\begin{align}
\partial_t \rho_{n} =& \sum_{m\ne n}^M \gamma_{nm}\rho_{m} - \sum_{m \ne n}^M \gamma_{mn} \rho_{n} \notag \\ 
&- \frac{1}{\hbar} \sum_j^{N_T} \xi_{n,j} \left(\mathbf{p}_j^- - \mathbf{p}_j^+ \right) \cdot \mathbf{E}. \label{diagEqn}
\end{align}
Here $\gamma_{nm}$ represents either the non-radiative decay rate between a higher level $m$ and a lower level $n$ 
or pumping rate from lower level $m$ to higher level $n$, $\rho_n \equiv N(\mathbf{x}) \rho_{nn}$ is
the total number of electrons in level $n$ of atoms at location $\mathbf{x}$. 
$\xi_{n,j}$ represents the relationship between the population of level, $n$ and a given lasing
transition, $j$; $\xi_{n,j} = 1$ if $n$ is the upper level of the transition, $\xi_{n,j} = -1$ if
$n$ is the lower level of the transition, and $\xi_{n,j} = 0$ if $n$ is not
involved in that lasing transition.

Thus the full set of semiclassical lasing equations are Eqs.\ (\ref{maxEqn}), (\ref{sepPolEqn}), (\ref{sepPolSolved}),
and (\ref{diagEqn}), which define the wave equation for the electric field, the total polarization in terms of the
constituent polarizations, the equation of motion for each constituent polarization, and the equation of motion
for the populations in each atomic level respectively. Since now the time evolution of the off-diagonal elements of the density 
matrix are contained in the polarization equations, we will henceforth represent the populations (diagonal elements) in terms
of a density {\it vector} with components $\rho_n(x)$.

\section{Solving the semiclassical lasing equations}

\subsection{Two-level gain medium \label{sec:MB-SALT}}

In the limit of a two-level gain medium with only a single lasing transition, 
Eqs.\ (\ref{sepPolEqn}), (\ref{sepPolSolved}), and (\ref{diagEqn}) are simplified to
\begin{align}
\partial_t \mathbf{P}_g^+(\mathbf{x},t) =& -\left(\gamma_{\perp} + i \omega_{a}\right) \mathbf{P}_g^+ - \frac{i N(\mathbf{x})d}{\hbar}\left(\boldsymbol{\theta} \cdot \mathbf{E} \right) \boldsymbol{\theta}^* \label{MB1} \\
\partial_t d(\mathbf{x},t) =& -\gamma_\parallel(d - d_0(\mathbf{x})) - \frac{2}{\hbar} \left(\mathbf{P}_g^- - \mathbf{P}_g^+ \right) \cdot \mathbf{E} \label{MB2}
\end{align}
where $\gamma_\parallel = (\gamma_{12} + \gamma_{21})$ is the relaxation rate of the inversion and 
\begin{equation}
d_0(\mathbf{x}) = \left(\frac{\gamma_{21} - \gamma_{12}}{\gamma_{21} + \gamma_{12}} \right)N(\mathbf{x})
\end{equation}
is the inversion in the absence of an electric field. These equations,  (\ref{MB1}), and (\ref{MB2}), along with
the wave equation, (\ref{maxEqn}),
comprise the Maxwell-Bloch equations, the most basic version of semiclassical laser theory.
As noted above, it is these equations which are solved directly for the steady-state properties using SALT.
For single mode lasing the SALT solution is essentially exact; for multimode lasing it requires the 
stationary inversion approximation (SIA), $\partial_t d = 0$.
This approximation holds if the multimode beating terms which drive time-dependence of the populations
are negligible.

The stationary inversion approximation (SIA), requires two conditions to be valid.
First, that the relaxation rate of the inversion, $\gamma_\parallel$,
be small compared to the modal spacing, $\Delta$, and
the dephasing rate of the polarization, $\gamma_\parallel \ll \Delta, \gamma_\perp$. 
This condition is usually met for microlasers, for example a Fabry-Perot laser cavity typically requires 
$L \le 100\mu$m for $\Delta \gg \gamma_\parallel$ \cite{haken91}.  The second
condition is that the mode spacing must be well separated from the relaxation oscillation frequency, so
as to avoid resonantly driving relaxation fluctuations which could destabilize the multimode
solution. The relaxation oscillation frequency is $\omega_r \sim \sqrt{\kappa \gamma_\parallel}$ \cite{arecchi_ABC_84}, where $\kappa$ is the field decay
rate of the cavity. For microcavities, $\kappa \le \Delta$, so $\omega_r \le \sqrt{\Delta \gamma_\parallel} < \Delta$.
As such, the SIA will hold and the multimode solution found by
SALT will be correct when $\gamma_\parallel \ll \Delta, \gamma_\perp$ \cite{ISALT}.  Essentially the same
conditions will be required for C-SALT to describe multimode lasing, although, as for SALT,
C-SALT solutions should be generality accurate for the single-mode case.

Continuing our brief review of SALT, we now simplify our analysis
by treating slab or two-dimensional geometries for which the electric fields
in the transverse magnetic (TM) modes,
can be taken to be a scalar, $\mathbf{E} \rightarrow E$), 
noting that the treatment discussed here is still completely
applicable in geometries for which the fields must be treated as vectors
\cite{esterhazy14}.  Having done so, we make a multimode ansatz, stating
that the electric field and polarization can be broken up into $N_L$ components
with distinct frequencies representing each lasing mode,
\begin{align}
E^+(x,t) &= \sum_\mu^{N_L} \Psi_\mu(x) e^{-i\omega_\mu t} \label{ansatz1} \\
P_g^+(x,t) &= \sum_\mu^{N_L} p_{\mu}(x) e^{-i\omega_\mu t} \label{ansatz2}
\end{align}
where the plus superscript denotes the positive frequency component
of the field, $E = 2 \textrm{Re}[E^+]$, and $\Psi_\mu(x)$ and $p_\mu(x)$ are the
spatial profiles of the electric field and corresponding polarization of the lasing
mode with frequency $\omega_\mu$. The multimode ansatz
allows us to match frequency components of the electric and polarization
fields through (\ref{MB1}),
\begin{equation}
p_{\mu} = \frac{|\theta|^2}{\hbar} \frac{d}{\omega_\mu - \omega_{a} + i\gamma_{\perp}}\Psi_\mu. \label{polSolvedEqn}
\end{equation}
Upon inserting this definition into (\ref{MB2}) and (\ref{maxEqn}) and making the rotating
wave approximation (RWA), which is well satisfied for all lasers of interest, one recovers the
coupled set of SALT equations.  As noted above, these take the form of a set of wave equations,
one for each lasing mode of the cavity, which are coupled
through the non-linear hole-burning interaction in the
gain medium \cite{tureci06,tureci07,ge10},
\begin{align}
0 =& \left[ \nabla^2 + \left(\varepsilon_c(x) + 4\pi \chi_g(x, \omega_\mu)\right)k_\mu^2 \right]\Psi_\mu(x), \label{saltfield} \\
\chi_g(x, \omega) =& \frac{|\theta|^2}{\hbar} \frac{ d_0(x)}{\omega - \omega_a + i \gper}  \notag \\
&\times \left( \frac{1}{1 + \frac{4 |\theta|^2}{\hbar^2 \gpar \gper} \sum_\nu^{N_L} \Gamma_\nu |\Psi_\nu|^2} \right), \label{saltchi}
\end{align}
where $\Gamma_\nu = \gper^2 / (\gper^2 + (\omega_\nu - \omega_a)^2)$ is the Lorentzian
gain curve evaluated at $\omega_\nu$. 

To solve the SALT equations, (\ref{saltfield}) and (\ref{saltchi}), simultaneously, each
of the lasing modes is expanded over a basis,
\begin{equation}
\Psi_\mu(x) = \sum_n a_n^{(\mu)} f_n(x; \omega_\mu)
\end{equation}
where the basis functions $f_n$ are fixed, (but possibly dependent upon the lasing
frequency $\omega_\mu$), and the expansion coefficients $a_n$ are found via a non-linear
solution algorithm. There are at least two useful choices for the basis set.  The first, and most developed
to this point, is to use a basis set of purely outgoing non-hermitian states which are termed
the Constant Flux (CF) states \cite{tureci06,ge10,li_thesis}.  More recently it has been shown that 
one can simply use a position space basis \cite{esterhazy14} combined with a perfectly matched layer (PML) 
to implement the outgoing boundary condition We will not go into the details of these computational
approaches here as either can be used with the C-SALT equations which are our main focus.  Instead we 
refer the interested reader to the cited references; the computations presented here are based on
the CF approach.

SALT assumes only a single lasing transition, and the above equations are then typically presented in
natural, scaled units for electric field and inversion, $E_c = 2 |\theta| / (\hbar \sqrt{\gamma_\perp \gamma_\parallel})$,
and $d_c = 4\pi |\theta|^2 / (\hbar \gamma_\perp)$, that involve the relaxation rates associated with that transition.
With multiple transitions with differing relaxation rates, all contributing to the lasing lines, there will be no
such natural scaling for C-SALT.  As noted above, previous work has shown that the steady-state equations
for N-level lasing with a single transition can be exactly mapped onto the SALT equations with renormalized relaxation parameters, 
so that computationally the N-level case is negligibly different from the two-level Maxwell-Bloch case \cite{cerjan12}.

\subsection{Multiple lasing transitions \label{Multiple lasing transitions}}

When multiple lasing transitions are present, the multimode ansatz is expanded
to include constituent polarizations,
\begin{equation}
p_j^+(x,t) = \sum_\mu^{N_L} p_{j,\mu}(x) e^{-i\omega_\mu t} \label{ansatz3}
\end{equation}
which still allows one to match frequency components for each
constituent polarization using Eq.\ (\ref{sepPolSolved}) to find
\begin{equation}
p_{j,\mu} = \frac{|\theta_j|^2}{\hbar}\frac{d_j}{\omega_\mu - \omega_{a,j} + i\gamma_{\perp,j}} \Psi_\mu. \label{chi1}
\end{equation}
To derive the C-SALT equations, this solution for the constituent polarization is inserted
into the equation of motion for the atomic levels, Eq.\ (\ref{diagEqn}), and the RWA is again
made, resulting in
\begin{align}
\partial_t \rho_n =& \sum_{m\ne n}^M \gamma_{nm}\rho_{m} - \sum_{m \ne n}^M \gamma_{mn} \rho_{n} \notag \\ 
&- \sum_j^{N_T}  \frac{2|\theta_j|^2 \xi_{n,j} d_j}
{\hbar^2\gamma_{\perp,j}}  \left(\sum_{\nu,\mu} \Gamma_{\nu,j}\Psi_\nu^* \Psi_\mu e^{-i(\omega_\mu - \omega_\nu)t} \right). \label{multTrans1}
\end{align}
To proceed, we first rewrite this equation in terms of a $M \times M$ population density vector
at each position in the cavity,
$\boldsymbol{\rho}(x)$, whose components are the atomic level populations $\rho_n(x)$.
Additionally, we make a slightly stronger version of the SIA, the stationary
population approximation (SPA), which states that the beating between
different lasing modes does not lead to significant time-dependence in the level populations, so
$\partial_t \boldsymbol{\rho} \approx 0$.
While a more detailed discussion of SPA will be given in section \ref{secspa}, it
is worth noting here that it is valid for most laser systems of interest,
i.e.\ those with fast decays into and out of the lasing levels, while the upper
lasing level of each transition is metastable and thus long lived. These fast
relaxation rates need not be small compared to $\Delta$ and $\gper$, only the
relaxation rates of the metastable upper lasing transitions need be slow in this sense, 
a condition which is 
is numerically tested and confirmed by the FDTD simulations below. 

Thus, the above equation
can be rewritten as
\begin{equation}
\mathbf{0}= R\boldsymbol{\rho} + \sum_{j}^{N_T} \frac{2|\theta_j|^2}
{\hbar^2\gamma_{\perp,j}}\left(\sum_{\nu} \Gamma_{\nu,j}|\Psi_\nu|^2 \right) \Xi_j \boldsymbol{\rho}, \label{ratemat}
\end{equation}
where $R$ is a matrix containing information about the pump and decay rates, and 
$\Xi_j$ is a matrix containing information about which level populations are coupled 
to the partial polarizations and constitute the $j$th lasing transition. The full forms of these matrices
are given in Appendix \ref{app:A}.

Eq.\ (\ref{ratemat}) is a homogeneous equation satisfied by the ``vector" of atomic populations
at each point in space. It requires knowledge of the lasing modes to be solved and hence will need
to be solved simultaneously with the electric field equations. In addition, 
in the absence of gain diffusion, the total number of gain atoms at each
point in space, $N(x)$, is externally fixed, and is a given of the problem. Hence the homogeneous
eq. (\ref{ratemat}) is to be solved subject to the normalization condition
\begin{equation}
\sum_n^M \rho_n(x) = N(x), \label{cond1}
\end{equation}
which uniquely determines the level population vector. 

It is convenient to incorporate this normalization condition directly into Eq.\ (\ref{ratemat})
by defining a matrix $B$ and an M-component total number vector $\mathbf{N}(x)$ such that
\begin{equation}
B \boldsymbol{\rho} = \mathbf{N}(x),
\end{equation}
where neither the matrix $B$, nor the vector $\mathbf{N}(x)$ are uniquely
defined, but must be chosen to represent Eq.\ (\ref{cond1}). The
normalization can then be inserted into Eq.\ (\ref{ratemat}),
resulting in
\begin{equation}
\boldsymbol{\rho}= \left[ R + B + \sum_{j}^{N_T} \frac{2|\theta_j|^2}
{\hbar^2\gamma_{\perp,j}}\left(\sum_{\nu}^{N_L} \Gamma_{\nu,j}|\Psi_\nu|^2 \right) \Xi_j\right]^{-1} \mathbf{N}(x). \label{popinveqn}
\end{equation}

To incorporate multiple lasing transitions into the
SALT formalism to recover the C-SALT equations, all that must be altered is the equation for the
electric susceptibility (\ref{saltchi}), not the wave equation itself.
Using (\ref{chi1}), the electric susceptibility can be written
as
\begin{equation}
  \chi_g(x,\omega) = \frac{1}{\hbar} \sum_j^{N_T} \frac{|\theta_j|^2 d_j}{\omega - \omega_{a,j} + i \gamma_{\perp,j}},
  \label{popchi}
\end{equation}
where $d_j \equiv \rho_n^{(j)} - \rho_m^{(j)}$ is determined from Eq.~(\ref{popinveqn}).
The main difference here is now the atomic population densities
cannot be directly inserted into the wave equation through the use of a scalar
inversion equation. Instead we must simultaneously solve the wave equation
(\ref{saltfield}), and the population equation (\ref{popinveqn}).
Using a non-linear iteration algorithm to solve the problem numerically, 
one inserts an initial guess for the field profiles, $\{ \Psi_\mu(x) \}$,
and uses these to solve for the full spatial profile of the atomic
population densities. This result for the population densities generates a 
guess for the susceptibility, which can be inserted into the wave equation
and iterated back and forth to self-consistency.

There is one other important difference between (\ref{saltchi}) and
(\ref{popchi}) in the case of a ``partially pumped''
laser, in which pumping is applied to only a subset of the regions
containing gain media \cite{andreasen_creation_2009,andreasen_effects_2010,ge_unconventional_2011,bachelard_taming_2012,hisch_pump-controlled_2013}.
One needs to distinguish physically between two situations: 1) Having a (given) spatial density of gain atoms, which can be non-uniform, and hence will lead to non-uniform but always positive gain under conditions of
uniform spatial pumping. This is taken into account from the specification of $N(x)$ under the assumed 
conditions of spatially uniform pumping.  2) Spatial non-uniform pumping ``partial pumping") in which 
there is a uniform distribution of gain atoms, which are not equally pumped.  This would appear in our
formalism as a variation with spatial position of the pumping rate parameters in the $R$ matrix of (\ref{popinveqn}). In this case non-pumped regions would act as absorbers for laser light, which would 
automatically be taken into account in terms of a spatial variation in the susceptibility function, $\chi_g (x)$, which would now have an absorbing form in those unpumped regions. 

\subsection{Gain diffusion}

The formalism developed in
Section \ref{Multiple lasing transitions} can be extended to include gain diffusion,
a phenomenon found in many types of gain media. A term
representing this effect can be added to Eq.\ (\ref{multTrans1}),
resulting in
\begin{align}
\partial_t \rho_n =& \sum_{m\ne n}^M \gamma_{nm}\rho_{m} - \sum_{m \ne n}^M \gamma_{mn} \rho_{n} + D_n \nabla^2 \rho_n \notag \\ 
&- \sum_j^{N_T}  \frac{2|\theta_j|^2 \xi_{n,j} d_j}
{\hbar^2\gamma_{\perp,j}}  \left(\sum_{\nu,\mu} \Gamma_{\nu,j}\Psi_\nu^* \Psi_\mu e^{-i(\omega_\mu - \omega_\nu)t} \right), \label{diffTrans1}
\end{align}
where $D_n$ is the longitudinal diffusion coefficient for the atomic level $|n\rangle$.
Despite the similarities of Eqs.\ (\ref{multTrans1}) and (\ref{diffTrans1}), there is one
important difference, namely that the atomic populations at each spatial location are
now coupled together. Thus, when diffusion is present, the population density
vector $\boldsymbol{\rho}$ is an $M \times P$ dimensional vector whose components are atomic populations
at each of $P$ discretized spatial locations.
The stationary population approximation can still be made, resulting in a generalized homogeneous
equation for the population density vector,
\begin{equation}
\mathbf{0}= \left(R + D\nabla^2 \right)\boldsymbol{\rho} + \sum_{j}^{N_T} \frac{2|\theta_j|^2}
{\hbar^2\gamma_{\perp,j}}\left(\sum_{\nu}^{N_L} \Gamma_{\nu,j}|\Psi_\nu|^2 \right) \Xi_j \boldsymbol{\rho}, \label{diffTrans2}
\end{equation}
where $D$ is the matrix of longitudinal diffusion coefficients at each spatial location, and the other two
matrices $R$, and $\Xi$ have also now been similarly expanded over the position basis as well.
To correctly normalize Eq.\ (\ref{diffTrans2}), we note that one consequence of SPA
is $\partial_t \sum_n \rho_n = 0$, and when performing this sum on Eq.\ (\ref{diffTrans1}),
one finds that the total population density is homogeneous in the steady state
in the presence of diffusion,
\begin{equation}
0 = \nabla^2 \left( \sum_n^M D_n \rho_n \right). \label{popdiffeqn}
\end{equation}
Furthermore, the walls of the cavity prevent any flux of gain atoms
across its borders, which is represented by the Neumann boundary condition
\begin{equation}
\partial_x \rho_n|_{x = 0,L} = 0.
\end{equation}
This, together with Eq.~(\ref{popdiffeqn}), yields the normalization
\begin{equation}
\sum_n^M \rho_n(x) = N.
\end{equation}
This is a simple restatement of the fact that a diffusive gain medium in a passive
cavity will be evenly distributed in the steady state, unlike a non-diffusive
gain medium where the density of gain atoms can fluctuate depending on their
distribution. 

Note however, that while the diffusion condition implies that the number
of ``atoms" will be uniform in space, their excitation distribution will not be.  Spatial hole-burning
due to the spatial variation of the lasing modes above threshold will lead to a non-uniform 
gain even in the presence of uniform pumping and diffusion, and this effect is still captured by the theory.
We will see below that these two effects compete; as the diffusion coefficient of the medium is increased, the
excitation distribution will become more uniform in steady-state, counteracting the effects of 
spatial hole-burning.

Again, the normalization requirement can now be expressed as
\begin{equation}
B \boldsymbol{\rho} = \mathbf{N},
\label{spatialnorm}
\end{equation}
where the matrix $B$ and vector $\mathbf{N}$ are
likewise expanded over the spatial basis.  Inserting
Eq.~(\ref{spatialnorm}) into Eq.\ (\ref{diffTrans2}) yields the
generalized level population equation
\begin{align}
\boldsymbol{\rho} =& \left[ R + D \nabla^2 + B \right.  \notag \\ 
& \left. + \sum_{j}^{N_T} \frac{2|\theta_j|^2}
{\hbar^2\gamma_{\perp,j}}\left(\sum_{\nu}^{N_L} \Gamma_{\nu,j}|\Psi_\nu|^2 \right) \Xi_j\right]^{-1} \mathbf{N}(x). \label{diffTrans3}
\end{align}
The solution now proceeds in the same way as in
Section \ref{Multiple lasing transitions}. Knowing the atomic level
populations $\boldsymbol{\rho}$, we can use Eq.\ (\ref{popchi}) to
solve for the susceptibility, which is then inserted into the wave
equation, Eq.~(\ref{saltfield}).  The problem thus reduces to a set of
differential equations, one per lasing mode, coupled through the
generalized level population equation (\ref{diffTrans3}).  The latter is
now a much bigger matrix equation, $MP \times MP$, as opposed to 
$M \times M$, increasing the computational cost, but still keeping it
within a manageable range, even for two-dimensional lasers (see below).

These two generalizations, Eqs.\ (\ref{popinveqn}) and
(\ref{diffTrans3}), along with their coupling to SALT,
Eqs.\ (\ref{popchi}) and (\ref{saltfield}), are the main results of
this paper.  They extend the capabilities of SALT to cover many
types of gain media, including newly-developed ones \cite{chua11,huang13}.
C-SALT can thus be used as an efficient tool for the
design and study of devices in which gain diffusion is
present alongside spatial hole-burning, a regime which is
challenging for traditional numerical methods to handle
\cite{jungo04,pernice07,hess08}.

\section{Results}

\begin{figure}[t!]
\centering
\includegraphics[width=0.45\textwidth]{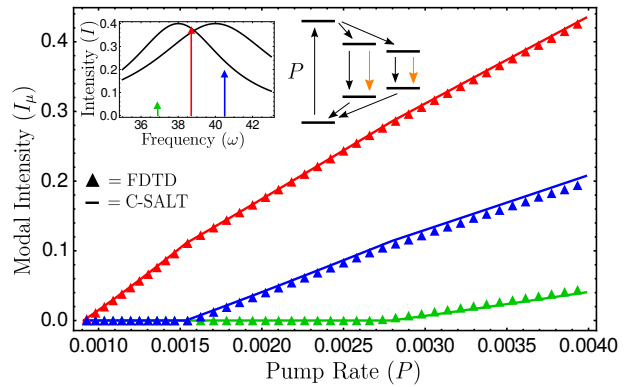}
\caption{Plot of modal intensities as a function of pump strength for a cavity
with $n=1.5$ and a gain medium consisting of atoms with two different atomic
transitions, $\omega_{a,1} = 40$, $\gamma_{\perp,1} = 4$, $\theta_1 = .1$, $\omega_{a,2} = 38$, $\gamma_{\perp,2} = 3$, $\theta_2 = .1$, and $6$ atomic levels in total,
with decay rates as indicated in the schematic. Results from C-SALT using the
stationary population approximation are shown as straight lines, results from FDTD
simulations are shown as triangles. The different colors indicate different lasing modes. 
Inset shows the modal frequencies and their intensities at $P = .0035$. All
values are reported in units of $c/L$. \label{twoplot}}
\end{figure}

To perform a well controlled test of the stationary population
approximation in C-SALT, for the case of multiple transitions without diffusion, 
we studied 1D microcavity lasers for which FDTD
simulations (without diffusion) are tractable.  We used a FDTD scheme similar to the one
proposed by Bid\'{e}garay \cite{bidegaray03}, altered to include
multiple lasing transitions and additional atomic populations. The
simulations were run for a total duration at least $40$ times greater
than the longest time scale in the model, to ensure a steady state was
reached. The simulated laser cavity consists of a
dielectric slab of background refractive index $n=1.5$, with a
perfectly-reflecting mirror on one side and an interface with air on
the other.  Distributed uniformly within the slab is a six-level
gain medium, with two atomic transitions of slightly different
frequencies and widths. For these set of simulations, the
diffusion constant was set to zero in the C-SALT equations.
The gain linewidths of the
transitions overlap, as shown in the inset of Fig.\ \ref{twoplot},
so each lasing mode receives significant gain contributions from
both transitions.
As shown in Fig.\ \ref{twoplot}, excellent
agreement is seen between C-SALT and FDTD simulations, both in predictions
of modal intensity and frequency, thus quantitatively verifying the use
of SPA. In these simulations, only the relaxation rates of the metastable
upper lasing levels are small compared to $\Delta$ and $\gamma_{\perp,j}$; the relaxation
rates of other atomic levels are of the same order.  This provides evidence for our
earlier claim that the only rates which must be small when compared to $\Delta$ and
$\gamma_{\perp,j}$ are those of the metastable upper lasing state.

\begin{figure}[t!]
\centering
\includegraphics[width=.45\textwidth]{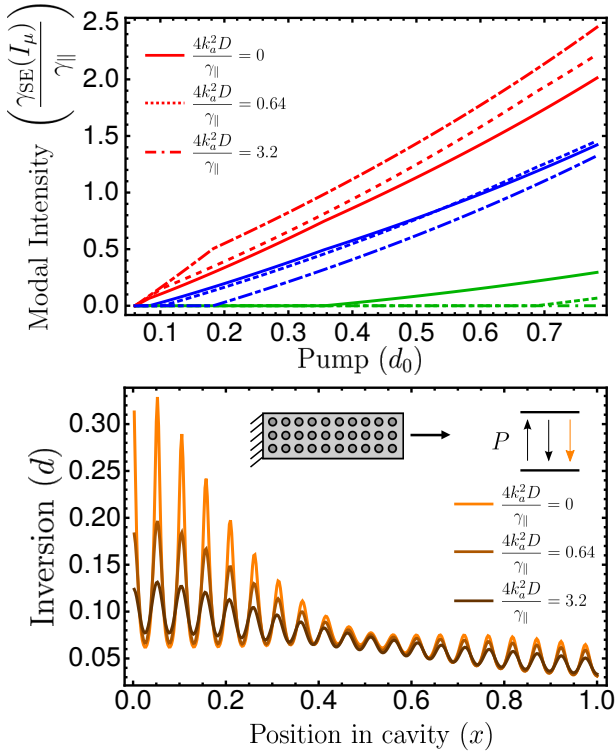}
\caption{(Top panel) Plot of the modal intensities calculated using C-SALT as a function of pump
strength for a dielectric slab cavity, $n=1.5$, and a two level, single transition
atomic gain medium, with $\omega_a = 40$ and $\gamma_\perp = 4$, values again in units of $c/L$. Simulations for three different diffusion strengths are shown in
solid, dashed, and dot-dashed lines. Different colors correspond to different lasing
modes within each simulation. (Bottom panel) Plot of the
inversion in the cavity as a function of position in the cavity at a pump strength
of $d_0 = 0.345$. 
Darker colors indicate increasing values of the diffusion
coefficient. Schematic shows a one-sided dielectric slab cavity containing a 
two level atomic medium subject to uniform
pumping.\label{diffplot}}
\end{figure}

We now move to the case of lasers with multimode lasing and gain diffusion, 
for which FDTD is a very challenging
computational effort, but which are relatively tractable using the C-SALT approach.
In Fig.\ \ref{diffplot}, we demonstrate how gain diffusion affects the
transition from gain-clamped single-mode lasing (which can be
described by the simplest form of laser theory \cite{haken}) to multimode lasing
(which is possible as a result of spatial hole-burning). The top panel of
Fig.\ \ref{diffplot}
shows C-SALT simulations for a two level atomic medium with three different values
for the diffusion coefficient. 
The solid lines show the modal intensities
of the medium without diffusion, and dotted and dot-dashed lines of the same color
show the evolution of the modal intensities as the diffusion coefficient is increased.
We observe, as expected, that increasing the diffusion coefficient
postpones the transition from single-mode to multimode operation,
by increasing the threshold of the second and higher-order lasing
modes. In fact, for the largest diffusion coefficient the third
lasing mode does not reach threshold within the pump values that were simulated. 
The bottom panel of Fig.\ \ref{diffplot} shows the inversion
of the gain medium inside the cavity as a function of position within the cavity
for a given pump value, $d_0 = 0.345$, a point at which all three simulations have
exactly two lasing modes on. The faster the diffusion, the more uniform
is the inversion in the presence of lasing.
Darker colors indicate increasing values of the diffusion coefficient.

The results in Fig.\ \ref{diffplot} make intuitive sense: an increased
diffusion coefficient spatially ``smooths out'' the population
inversion of the first lasing mode; this supplies the first mode with
more gain, acting against the effects of spatial hole-burning.
Specifically, the bottom panel of Fig.\ \ref{diffplot} shows that
increasing the diffusion coefficient flattens the inversion close to
the cavity mirror, where the lasing modes spatially overlap. Near the end facet of the cavity, the gain is already being
used fairly uniformly, so the effects of diffusion are substantially
reduced.
These results can also be understood quantitatively by writing
down an effective inversion
equation for an atomic gain medium, $d$, which can be done
as there is only a single lasing transition \cite{cerjan12},
\begin{equation}
\partial_t d = -\gpar(d-d_0(x)) + D \nabla^2 d - \frac{4|\theta|^2}{\hbar \gamma_\perp} \sum_\nu^{N_L} \Gamma_\nu |E_\nu|^2 d, \label{diffd}
\end{equation}
where $d_0(x)$ is the equilibrium value of the inversion density in the absence
of both fields inside the cavity and diffusion.
The final term on the right hand side can be identified as the rate of stimulated
emission, 
\begin{equation}
\gse(I) = \frac{4|\theta|^2}{\hbar \gamma_\perp} \sum_\nu^{N_L} \Gamma_\nu |E_\nu|^2,
\end{equation}
which is spatially dependent and proportional to the intensity
of the local electric field. Using the SPA, Eq.\ (\ref{diffd}) can be solved for the inversion density, as
\begin{equation}
d(x) = \left[1 + \frac{\gse(I)}{\gpar}- \frac{D}{\gpar} \nabla^2 \right]^{-1} d_0(x).
\label{dx}
\end{equation}
Thus, for diffusion to be germane to the system, 
\begin{equation}
\frac{k_D^2 D}{\gpar} \gtrsim 1 + \frac{\gse(I)}{\gpar}, \label{diffineq}
\end{equation}
where $k_D$ is the wavevector associated with the scale of the inhomogeneity
in the inversion. If the inversion is less than this, the gain atoms are unable to move
very far before they either non-radiatively decay or undergo stimulated emission, preventing
the diffusion from washing out the effects of the spatial inhomogeneity in the inversion.
For spatial hole-burning, this variation in the inversion is
on the order of the atomic transition wavelength, $k_D = 2k_a$, as the inversion oscillates twice as fast as the electric field \cite{gordon08}.
This prediction is consistent with the numerical results shown in Fig.\ \ref{diffplot}, where
the onset of strong diffusion suppresses multimode operation and leads to
gain clamped behavior.

\begin{figure}[t!]
\centering
\includegraphics[width=.45\textwidth]{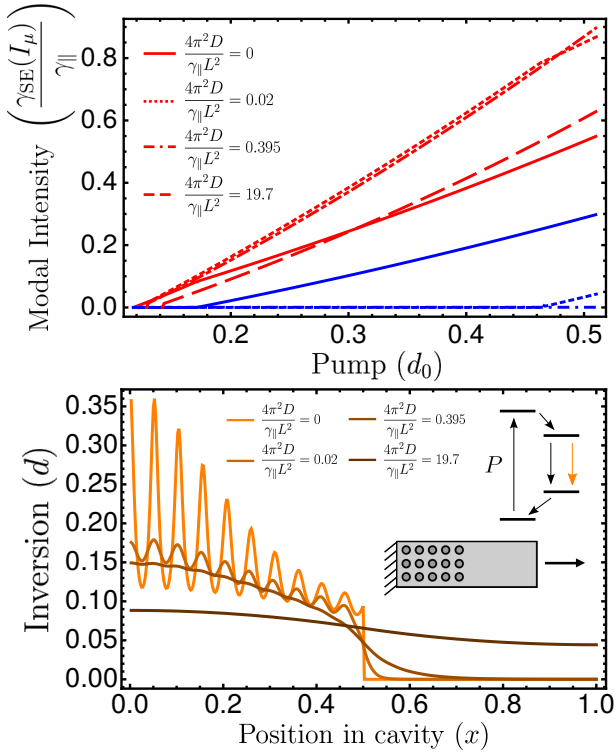}
\caption{(Top panel) Plot of the modal intensities calculated using SALT as a function of pump
strength for a partially pumped dielectric slab cavity, $n=1.5$, containing a four level, single transition
atomic gain medium. Simulations of four different values of diffusion are shown as solid, dotted, dot-dashed, and dashed lines.
The first lasing mode to turn on in all of the simulations is shown in red, and the second lasing mode,
which only turns on in two of the simulations, in blue. (Bottom panel) Plot of the
inversion in the cavity as a function of position in the cavity at a pump strength
of $d_0 = 0.37$. 
Darker colors indicate increasing values of the diffusion
coefficient. Schematic shows a partially pumped one-sided dielectric slab cavity
containing a four level atomic medium with a single lasing transition.\label{partplot}}
\end{figure}

For partially pumped cavities \cite{andreasen_creation_2009,andreasen_effects_2010,ge_unconventional_2011,bachelard_taming_2012,hisch_pump-controlled_2013}, 
there is, in addition to the scale associated with smoothing out spatial hole-burning, 
another relevant scale for measuring the strength of diffusion.
This is the scale at which the diffusion
begins to overcome the spatially inhomogeneous pumping, an inhomogeneity on the scale of the
entire cavity length.
Quantitatively, this can still be understood from Eq.\ (\ref{diffineq}), except that now
$k_D = 2\pi/L$, where $L$ is the length of the cavity, and we are assuming that the variation
of the pumping is on the scale of the entire cavity. This leads to the 
criterion that when $4\pi^2 D/L^2 > \gamma_\parallel + \gse(I)$ diffusion will strongly reduce the
effects of non-uniform pumping by allowing the inversion to penetrate into the non-pumped regions.

\begin{figure}
\centering
\includegraphics[width=.45\textwidth]{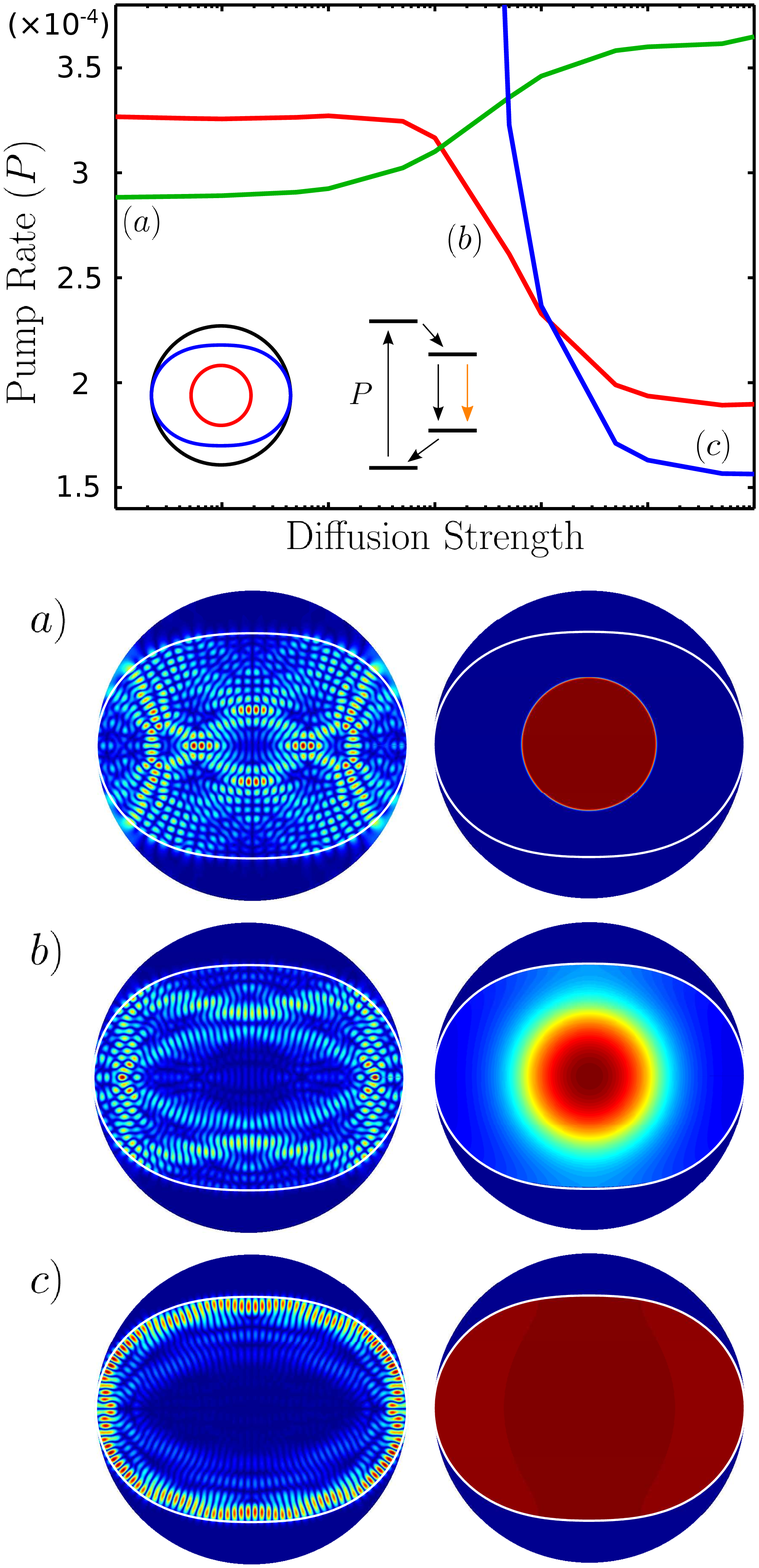}
\caption{Plot of the non-interacting modal thresholds as a function of the diffusion
coefficient in a quadrupole cavity with $\epsilon = 0.16$, $r_0 = 3.45\mu m$, $k=6.27 \mu m^{-1}$, 
$\gamma_\perp = .19\mu m^{-1}$, $n=3 + .004i$. Only the three modes that become the threshold lasing
mode for different values of the diffusion are plotted. Left schematic within the plot shows the boundary of the simulated
region, black circle, boundary of the cavity, blue quadrupole, and applied pump profile, red circle. 
The right schematic within the plot depicts the four level gain medium that fills the entire quadrupole cavity. Markers
in the plot correspond to the threshold lasing mode and corresponding inversion profile accounting for
the effects of diffusion at their locations. The inversion profile is shown in a false color plot, 
with red corresponding to large inversion, and blue corresponding to small inversion. 
The white boundary in the plots denotes the cavity boundary. \label{quadplot}}
\end{figure}

Both types of diffusion-induced transitions are demonstrated
in Fig.\ \ref{partplot}, which shows the results of C-SALT simulations
of a four-level gain medium with a single transition, which is only
pumped in half the cavity. A two-level medium would not be suitable
for this test as it would have strong absorption in the unpumped region.
As the diffusion coefficient is increased, the first transition from
the spatial hole-burning regime to the spatially-averaged gain
saturation regime is observed, as well as the effects of gain clamping
which increases the second lasing threshold.  In this regime, the
inversion does not penetrate into the unpumped region. However, as the
diffusion coefficient is further increased, the inverted atoms are able
to penetrate further into the unpumped region.  For this example, the
effective spatial scale of the partial pumping is the length of the
cavity, $L$, by choice. As expected, we observe this second transition
when $ 4 \pi^2 D / L^2 \gpar \sim 1 + \gse(I)/\gpar$. 
Once the first transition has occurred the gain is sufficiently clamped, even though
it is not uniform over the entire cavity, and we only find single mode lasing.
If one intentionally pumps non-uniformly on a smaller scale than $L$, that would
decrease the diffusion coefficient needed for the second transition until in the
case of wavelength scale non-uniform pumping the two transitions would coincide
roughly.


To demonstrate the scalability of C-SALT to multiple dimensions we ran
2D TM simulations of quadrupole-shaped dielectric cavities, whose boundary is defined
by the equation $r(\theta) = r_0(1 + \epsilon \cos(2\theta))$, where
$\epsilon$ represents the degree of deformation from a circular cavity.
Such cavities have been extensively studied in the context of wave chaos theory and
experiments \cite{nockel97,gmachl98,li_thesis}. It was found that the spatial profile of the first lasing
mode depends upon both the deformation and the pumping profile \cite{nockel97,gmachl98,li_thesis}.
Here we show in Fig.~\ref{quadplot} that for a quadrupole cavity with $\epsilon = 0.16$
which is being partially pumped in the middle of the cavity, the spatial profile
of the threshold lasing mode changes with the strength of the carrier diffusion
in the system. When the diffusion strength is too weak to overcome the partial
pumping of the cavity, the first threshold mode is found to have strong angular
dependence in its far-field intensity output and heavily overlap the center of the
cavity where the gain medium is inverted. As the diffusion coefficient is increased,
the threshold lasing mode changes to become one that lives closer to the edge of the
cavity, increasing its lifetime. Finally, when the gain becomes nearly uniform due
to diffusion, we find that the threshold lasing mode is a whispering gallery mode.

\section{Discussion of SPA \label{secspa}}

In the derivation of the population equation given above, (\ref{popinveqn}),
we took as an assumption that the beating terms from the coupling of different
mode amplitudes averaged to zero and thus could be neglected. In this section
we will make explicit this approximation. Similar to the discussion for a two-level
gain medium in section \ref{sec:MB-SALT}, there are two criteria that go into
the SPA: the populations do not acquire beating terms in the presence of multiple
lasing modes, and that relaxation oscillations are not resonantly enhanced by being driven at the
beat frequency. As discussed above, for a two level medium, the former criterion
requires that $\gamma_\parallel \ll \Delta, \gamma_\perp$, and the later requirement
that $\omega_r \sim \sqrt{\kappa \gamma_\parallel} < \Delta$. The main difficulty
encountered when generalizing these equations in the presence of multiple transitions
is the absence of an explicit formula for an effective $\gamma_\parallel$ parameter entering these
inequalities. However, the example of the N-level single transition case \cite{cerjan12}, for which
we have such an explicit formula, strongly suggests if all of the 
$ \{\gamma_{lu} \}$ are small in the relevant sense, (where
$\{\gamma_{lu}\}$ is the set of non-radiative decay rates from the upper lasing level
to the lower lasing level of each lasing transition) then the SPA will hold. Thus,
since $\kappa \leq \Delta$ (they are comparable for most of the lasers studied here),
as long as $\sqrt{\Delta} > \sqrt{ \gamma_{lu}^j}, \forall_j$, we expect that the SPA will be satisfied.  
This is consistent with our FDTD 
simulation results, as noted above.

In the limit of small violations of the SPA, it is possible to correct perturbatively for the
effects of the beating populations within a generalized SALT framework.
This calculation is discussed in Appendix \ref{app:B}.

\section{Semi-SALT: a free-carrier model \label{semiSec}}

A natural extension of the discussion on atomic gain media with
multiple transitions is to a treatment of bulk semiconductor gain
media, in which there is a continuum of available transitions for the
electrons between the conduction and valence
bands. In doing so, we must also consider Pauli exclusion
and Fermi-Dirac statistics. The polarization-Bloch equation for
semiconductor media was originally derived by Lindberg and Koch, who took into 
account many-body effects \cite{lindberg88},
\begin{multline}
(\hbar \omega - \Delta \varepsilon_q + i \hbar \gamma_q ) \rho_{cv,q}(r,\omega) \\
= \left(f_{c,q}(r) - f_{v,q}(r) \right) \Big[g_q E(r,\omega) \\
 + \frac{1}{V} 
\sum_{q'} V_s(q-q') \rho_{cv,q'}(r,\omega) \Big], \label{semipoleqn}
\end{multline}
in which $\rho_{cv,q}$ is the off-diagonal density matrix element between
the conduction and valence bands at electron momentum $q$,
$g_q$ is the dipole matrix element for the transition at $q$, $\Delta \varepsilon_q$ is
the re-normalized energy difference between the conduction and valence states,
$\gamma_q$ is the dephasing rate, and $V_s$ is the Coulomb interaction. 
Note that the inversion term is
the simplification of $f_c(1-f_v) - f_v(1-f_c)$, the probability that
a conduction state is filled and the relevant valence state is open minus
the probability that a valence state is filled and the conduction state
is open. The macroscopic polarization field in Maxwell's equations
is then given by \cite{koch89}
\begin{equation}
P(r,\omega) = \frac{1}{V} \sum_q g_q \rho_{cv,q}(r,\omega).
\end{equation}

To proceed we will make the free-carrier approximation, setting
$V_s = 0$, with full understanding that Coulomb repulsion is an
important effect in semiconductor lasers, and thus that the results found
here are just the first step towards a more complete theory. Doing so allows one to write
down the free-carrier susceptibility,
\begin{align}
\chi(r,\omega) = 
\int d^3 q \frac{2}{(2 \pi)^3} g_q^2 \frac{f_{c,q}(r;\phi,|E|) - f_{v,q}(r;\phi,|E|)}
{\hbar \omega - \Delta \varepsilon_q + i \hbar \gamma_q},
\end{align}
where the factor of two appearing in the numerator accounts for
spin degeneracy. The factors of $f_{c,q}$ and $f_{v,q}$ are 
simply the occupation probabilities for finding an electron at momentum $q$ in the
conduction and valence bands respectively and depend both upon the applied electric potential,
$\phi$, as well as the magnitude of the electric field within the cavity. The inversion equations, 
taking into account
Fermi-Dirac statistics, take the form:
\begin{align}
\partial_t D_q(r) =& -\gamma_{\parallel,q}(D_q(r) - D_q^{(0)}) \notag \\
&- \frac{2}{i \hbar}((E g_q^* \rho_{cv,q})^* - c.c.) \\
D_q(r) =& f_{c,q}(r) - f_{v,q}(r) \\
D_q^{(0)} =& f_{c,q}^{(0)} - f_{v,q}^{(0)} \notag \\
=& \frac{1}{e^{\beta(\varepsilon_{c,q} - \mu - e\phi)} + 1} 
- \frac{1} {e^{\beta(\varepsilon_{v,q} - \mu + e\phi)} + 1},
\end{align}
where $\gamma_{\parallel,q}$ is the non-radiative interband relaxation rate
between the conduction and valence bands at momentum state $q$, and 
$f_{c,q}^{(0)}$ is the Fermi function for the conduction band at momentum $q$
in which we have introduced the electro-chemical potential $\mu + e \phi$.
In this theory the applied voltage $\phi$, related to the injected current,
will play the role of the pump. In writing such a simple inversion equation
we are neglecting intraband transitions, which is why a population equation
similar to Eq.\ (\ref{popinveqn}) is not needed here.

Next, we expand $E(r,t)$ and $\rho_{cv,q}(r,t)$
as a summation of distinct lasing modes with different spatial profiles in
the same manner as done previously (\ref{ansatz1}), (\ref{ansatz2}). 
Using the above equations, we again assume stationary inversion of all of the constituent transitions
$\partial_t D_q(r) = 0$, assuming that the modal beating terms in the product $E \rho_{cv,q}^*$ are
negligible, so that we can solve for the susceptibility and insert it into
Maxwell's wave equation, resulting in the Semi-SALT equations:
\begin{align}
[ \nabla^2 &+ \left( \varepsilon_c + 4\pi \chi_g(r,\omega)
 \right) k_\mu^2 ] \Psi_\mu = 0 \label{semi1} \\
\chi_g(r,\omega) =& \int d^3q \frac{2}{(2 \pi)^3} g_q^2 \frac{f_{c,q}^{(0)} - f_{v,q}^{(0)}}
{\hbar \omega - \Delta \varepsilon_q + i \hbar \gamma_q} \notag \\
&\times \left(\frac{1}{1+ \frac{4 g_q^2}{\gamma_\parallel \hbar} \sum_\nu \Gamma_{\nu,q}
|\Psi_\nu(r)|^2} \right) \label{semi2}
\end{align}
where
\begin{equation}
\Gamma_{\nu,q} = \frac{\hbar \gamma_q}{(\hbar \omega_\nu - \Delta \varepsilon_q)^2 + 
\hbar^2 \gamma_q^2}
\end{equation}
is the Lorentzian linewidth.

To solve the Semi-SALT equations we will assume that we are modeling a 
direct band-gap semiconductor laser where the renormalized energy gap can
be written as 
\begin{equation}
\Delta \varepsilon_q = \frac{\hbar^2 q^2}{2 m_r} + E_g
\end{equation}
in which $m_r$ is the reduced mass and $E_g$ is the energy gap when $q=0$. 
If we take both $\gamma_q$ and $g_q$ to be independent of $q$, we find that
the integral defining the real part of $\chi$ is divergent, as both the numerator and the denominator
$\propto q^4$, and for large $q$, $D_q^{(0)} \rightarrow -1$. This is a
known problem \cite{koch}, which stems from the fact that the Lorentzian line-shape
of the dipole transition is too broad and that other many-body effects
truncate the line-shape. However, within the assumptions already used in this
treatment the integral can be regularized by incorporating the correct $q$ dependence in to $g_q$ \cite{koch},
\begin{equation}
g_q = \frac{g_0}{1 + \frac{\hbar^2 q^2}{2 m_r} \frac{1}{E_g}}
\end{equation}
in which
\begin{equation}
g_0 = \langle \lambda' | \hat p | \lambda \rangle
\end{equation}
and where $| \lambda \rangle$ represents a lattice periodic function, $\hat p$ is the
momentum operator, and as such $g_0$ has no further dependence upon $q$.
Note that even when many-body effects are considered,
$\gamma_q$ is still relatively constant as a function of $q$.

Numerically solving the Semi-SALT equations is substantially more computationally demanding than the usual
SALT \cite{cerjan12}, or C-SALT equations discussed above.
The reason for this is two-fold.  First, the electric susceptibility no longer
depends linearly upon the pump variable. Instead the applied electric potential,
$\phi$, appears in the Fermi-Dirac functions defining the equilibrium inversion
of the semiconductor, yielding a nonlinear dependence in the susceptibility on the applied potential. This
has the effect of making the lasing threshold problem non-linear in the pump variable and thus increasing
the computational difficulty, both
to find the first lasing threshold, and to find subsequent laser thresholds for
additional modes. Second, the integral over $q$ must
be performed at each spatial location, another computationally expensive task. 
Both of these difficulties can be surmounted;
the details can be found in Appendix \ref{app:C}.

\begin{figure}
\centering
\includegraphics[width=.45\textwidth]{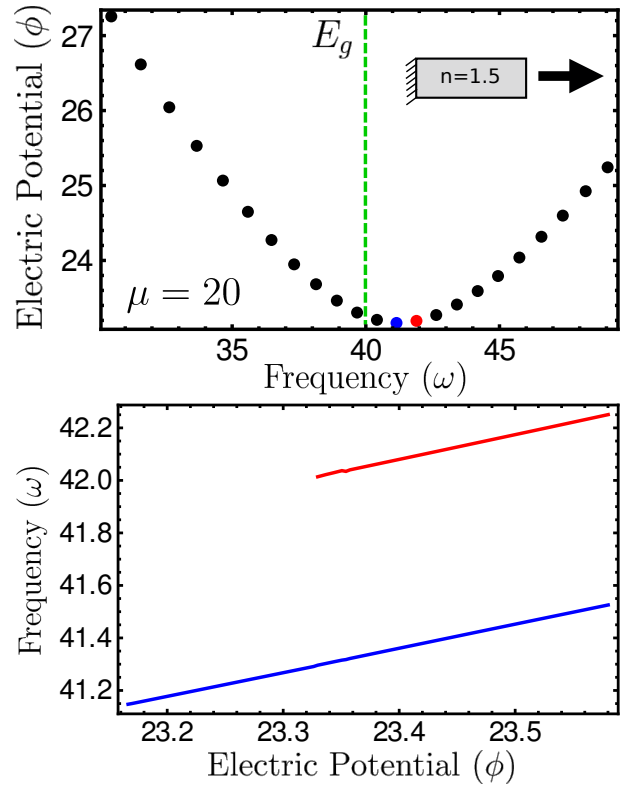}
\caption{(Top panel) Plot of the lasing thresholds for each of the non-interacting
modes in the cavity as a function of the applied electric potential, $\phi$, for
a single sided, slab semiconductor laser. The energy gap at $q=0$ has been set to
$E_g = 40$, and the chemical potential set at half that, $\mu=20$. The non-interacting
thresholds of the two modes studied in the bottom panel are given the same colors as appear
in the bottom panel. (Bottom panel) Plot of the frequencies of the two lasing modes
as a function of the applied electric potential. The frequency for the second lasing mode (red) 
is only shown after it reaches threshold. \label{semifig}}
\end{figure}

Even under the limiting assumptions made above, Semi-SALT is able
to predict two features unique to semiconductor lasers, as seen in Fig.~\ref{semifig}.
Unlike the atomic gain media discussed above, the gain curve of a semiconductor
gain media is in homogeneously broadened and asymmetric due to the continuum of transitions available above
the energy gap. As such, we expect that the thresholds of lasing modes with frequencies
above the energy gap will experience more gain and thus have lower thresholds, an
effect clearly seen in the top panel of Fig.~\ref{semifig}. The frequencies of
semiconductor laser modes are also expected to shift away from the energy gap as
the pump is increased due to Pauli blocking, additional electrons excited across
the band gap from the increasing the applied electric potential must find higher
energy states to transition to, yielding more energetic stimulated emission transitions.
This effect is clearly predicted by Semi-SALT, as shown in the bottom panel of Fig.~\ref{semifig}.

In general, quantitative computational study of semiconductor lasers is very challenging and
is only feasible with supercomputers and specialized codes.  We hope that incorporating the
steady-state ansatz and the SIA into the electromagnetic part of the calculation, as is done in 
Semi-SALT, could eventually lead to more efficient computational approaches to parts of this problem.

\section{Summary}

We believe that the C-SALT equations derived here are the most general and accurate form 
of the steady-state semiclassical lasing equations, assuming only stationary populations
and neglect of atomic and multi-level coherences, as is typically valid
for most lasers of interest.  As we have shown, these equations include both the 
effects of multiple lasing transitions and of gain diffusion and are computationally
tractable, even in more than one dimension. This will allow efficient simulation and
modeling of a number of new laser systems with these features.
The ability to model gain diffusion 
revealed two relevant scales for the diffusion constant.  Spatial hole burning disappears
and single mode lasing (gain clamping) sets in when $4 k_a^2 D \sim \gpar + \gse(I)$.
This could be demonstrated because C-SALT treats the above threshold gain competition
exactly. Non-uniform pumping effects disappear when $4 \pi^2 D / L^2 \sim \gamma_\parallel + \gse(I)$, where 
$L$ is the typical scale of the non-uniformity in the pump profile.  Comparisons with experiment
can be used to extract information about the relevant diffusion constant(s).
Finally, a free-carrier version of SALT, semi-SALT, was derived which fully incorporates
Fermi-Dirac statistics, showing the effect of Pauli blocking.  It is hoped that further work along
these lines will incorporate many-body effects leading to 
advances in semiconductor laser modeling.

\appendix
\section{Matrix definitions \label{app:A}}
The rate matrix, $R(x)$, is an $M \times M$ matrix where $M$ is the number of atomic levels,
and can be spatially dependent when partial pumping
is applied to the cavity. In Eq.\ \ref{ratemat}, $R$ can be calculated explicitly as
\begin{align}
R_{nn}(x) =& - \sum_m \gamma_{mn}(x) \\
R_{nm}(x) =& \gamma_{nm}(x), \; \; \; \; \; \; \; (n \ne m)
\end{align}
where in most cases the non-radiative decay rates $\gamma_{nm}$ are
a property of the atomic medium, and hence are not spatially dependent.
The coupling matrices $\Xi_j$ also have size $M \times M$ in Eq.\ \ref{ratemat}, and are defined as
\begin{align}
\Xi_{uu}^{(j)} &= \Xi_{ll}^{(j)} = -1 \\
\Xi_{ul}^{(j)} &= \Xi_{lu}^{(j)} = 1
\end{align}
where the indices $u$ and $l$ refer to the upper and lower
lasing levels of the constituent polarization $j$ respectively.

In Eqs.\ (\ref{diffTrans2}) and (\ref{diffTrans3}), these matrices have size
$PM \times PM$ where $P$ is the number of discretized spatial points in the cavity 
and $M$ is the number of atomic
levels. The definitions of these matrices are then adjusted to be zero everywhere except 
at the location $x \in P$, and have only a potentially non-zero block of size $M \times M$ at
this location in the matrix.

\section{Treatment of beating populations \label{app:B}}
We will now investigate the breakdown of the SPA when the beat frequencies are picked
up in the atomic population dynamics.
Following Ge \textit{et al.}\ \cite{ge08}, we assume that there are only two lasing modes in the cavity and
will only be concerned with the lowest frequency components such as $\omega_1 - \omega_2$,
neglecting the other side band terms that are generated such as $2\omega_1 - \omega_2$
as these will oscillate much faster. We first calculate corrections to the
atomic population densities defining
\begin{equation}
\boldsymbol{\rho}(x,t) = \boldsymbol{\rho}_s(x) + \left( \boldsymbol{\rho}_b(x,t) + c.c. \right),
\end{equation}
where the subscript $b$ denotes the portion of the population density beating at the difference frequency,
and $s$ denotes the steady state component.
Using this to rewrite (\ref{diffTrans1}) and separating out the time dependent portions we
find
\begin{align}
\mathbf{0} =& D \nabla^2 \boldsymbol{\rho}_s(x) + R \boldsymbol{\rho}_s(x) \notag \\
&+ \sum_{j=1} \frac{1}{i \hbar}\left(\Psi_1 p_{1,j}^* + \Psi_2 p_{2,j}^* - c.c.\right) \boldsymbol{\xi}_j, \label{eq3} \\
\partial_t \boldsymbol{\rho}_b(x,t) =& D\nabla^2 \boldsymbol{\rho}_b(x,t) + R \boldsymbol{\rho}_b(x,t) \notag \\
&+ \sum_{j=1} \frac{1}{i \hbar}\left(\Psi_1 p_{2,j}^* - \Psi_2^* p_{1,j}\right) e^{-i(\omega_1 - \omega_2)t} \boldsymbol{\xi}_j, \label{eq2}
\end{align}
where $\boldsymbol{\xi}_j$ is the vector form of the elements $\xi_{n,j}$.

As we are neglecting higher order beating frequencies, there two contributions to each
atomic polarization for the first mode, $p_{1,j}$, the first from the electric
field oscillating at $\omega_1$ coupled to the stationary population terms and a second
from the electric field oscillating at $\omega_2$ coupled to the beating polarization
at frequency $\omega_1 - \omega_2$.
\begin{equation}
p_{1,j} = \frac{g_j^2 \gamma_{1,j}}{\hbar}
\left(\Psi_1 d_{j,s} + \Psi_2 d_{j,\Delta} \right),
\end{equation}
where $d_{j,s}$ and $d_{j,\Delta}$ are the stationary and beating inversions associated with
lasing transition $j$ and
\begin{equation}
\gamma_{\mu,j} = \frac{1}{\omega_\mu - \omega_{a,j} + i \gamma_{\perp,j}},
\end{equation}
in which the time dependence has been separated from the beating population densities,
$\boldsymbol{\rho}_b = \boldsymbol{\rho}_\Delta e^{-i(\omega_1 - \omega_2)t}$. To leading
order the polarization of the $j$th transition can be written as
\begin{equation}
p_{\mu,j}^{(0)} = \frac{g_j^2 \gamma_{\mu,j}}{\hbar}
\Psi_\mu d_{j,s},
\end{equation}
and this is inserted into (\ref{eq2}) to obtain
\begin{align}
\boldsymbol{\rho}_\Delta =& \frac{1}{-i(\omega_1 - \omega_2)}\left( D \nabla^2 + R \right) \boldsymbol{\rho}_\Delta \notag \\ 
&+ \sum_j \frac{g_j^2}{\hbar^2} \left( \frac{\gamma_{2,j}^* - \gamma_{1,j}}{\omega_1 - \omega_2} \right)
\Psi_1 \Psi_2^* \Xi_j \boldsymbol{\rho}_s,
\end{align}
which can be inverted to find $\boldsymbol{\rho}_\Delta$,
\begin{equation}
\boldsymbol{\rho}_\Delta = M(\omega_1, \omega_2) \Psi_1 \Psi_2^* \boldsymbol{\rho}_s, \label{beateqn}
\end{equation}
where the matrix $M(\omega_1, \omega_2) = M(\omega_2, \omega_1)^*$ contains
all the information about the diffusion rates, decay rates, and field
frequencies. Using this, we are able to write down the first correction
to the atomic polarizations,
\begin{equation}
p_{1,j}^{(1)} = \frac{g_j^2 \gamma_{\mu,j}}{\hbar}
\Psi_1 \left(1 + M(\omega_1,\omega_2)|\Psi_2|^2 \right) d_{j,s}.
\end{equation}
Finally, this can be inserted into (\ref{eq3}) to calculate the corrections
to the steady
state population densities. However, it
is difficult to glean any analytic insight from these equations as there
is no obvious choice of parameters to compare, with multiple atomic
polarization relaxation rates and interlevel decay rates. Fortunately,
as mentioned above, the beating population densities are negligible in
the parameter regimes studied here, namely when the upper lasing levels of
the atomic transitions are metastable and have decay rates much less
than the atomic polarization decay rates and the free spectral range.

\section{Computation of Semi-SALT \label{app:C}}
There are two main new difficulties that one encounters upon implementing
Semi-SALT that are not seen in SALT computations using atomic gain media.
The first is that the lasing threshold problem is no longer linear in
the pump parameter. To find the non-interacting laser thresholds using the
TCF basis \cite{ge10} one must solve the equation,
\begin{equation}
\eta_n(\omega_\mu) = 4\pi \int \frac{2}{(2\pi)^3} d^3q g_q^2
\left(\frac{D_q^{(0)}}{\hbar \omega_\mu - \Delta \varepsilon_q + i \hbar \gamma_q} \right),
\end{equation}
which cannot be reformulated as a linear eigenvalue problem. Here,
$\eta_n(\omega)$ is the eigenvalue being tracked across frequency space
of the TCF basis equation for which we are attempting to determine where
the threshold lasing state it corresponds to reaches threshold. The TCF
basis is defined by the equations,
\begin{align}
0 =& [\nabla^2 + (\varepsilon_c(x) + \eta_n F(x)) \omega^2] u_n(x) \\
\partial_x u_n(L) =& ik u_n(L),
\end{align}
where $\eta_n$ is the eigenvalue for the $n$th solution of this
equation at frequency $\omega$, $u_n(x)$ is the eigenstate of
the $n$th solution of the TCF equation at $\omega$, and $F(x)$
is the pump profile.
Instead, 
to find all of the laser thresholds one must use both the real and imaginary
parts of this equation to first solve for $\phi_{thr}$ where $\re[\eta(\omega)] = 4\pi \re[\chi(\omega,\phi_{thr})]$,
and then solve for the offset in the imaginary part of this equation,
\begin{equation}
\delta = \im[\eta(\omega)] - 4\pi \im[\chi(\omega,\phi_{thr})].
\end{equation} 
By monitoring for when $\delta$ changes sign as the frequency
is swept through, the lasing threshold
can be found.

The second main computational problem arises from attempting to
directly solve the Semi-SALT equations, (\ref{semi1}),(\ref{semi2}), using
a non-linear solver above threshold. Such an algorithm for finding the
solution to the Semi-SALT equations requires evaluating the integrals
over the electron momentum at every point in space for every guess at
the lasing mode amplitudes and frequency until the non-linear solver converges, and 
is hopelessly inefficient. Instead, for each iteration of
the Semi-SALT equations in
the pump variable
above threshold this difficulty can be sidestepped by performing a
Taylor expansion on the electric susceptibility in terms of the
variables solved for at each iteration above threshold,
\begin{align}
\chi(\omega_{N+1}, \vec a_{N+1}) =& \chi(\omega_{N}, \vec a_{N}) +  \left. \frac{\partial \chi}{\partial \vec a}\right|_{N} \cdot (\vec a_{N+1} - \vec a_N) \notag \\
&+  \sum_\nu \left. \frac{\partial \chi}{\partial \omega_\nu} \right|_{N} (\omega_{\nu,N+1} - \omega_{\nu,N})
\end{align}
where $\chi(\omega_{N}, \vec a_{N})$ is the self-consistent solution for the electric
susceptibility for the applied electric potential $\phi_N$, and is a function
of all of the different lasing frequencies present, $\{ \omega_\mu \}$ and
the decomposition of the all of the lasing modes $\{ \Psi_\mu \}$, where,
\begin{equation}
\Psi_\mu(x) = \vec a^{(\mu)} \cdot \vec u(x; \omega_\mu),
\end{equation}
is the spatial decomposition of each lasing mode over the TCF basis at the
correct frequency. Using this Taylor expansion of the susceptibility, 
now all of the spatially dependent integrals can be evaluated before invoking
the non-linear solver, dramatically improving performance.

\begin{acknowledgments}
We thank Arthur Goestchy and Hui Cao for helpful discussions. We thank
the Yale High Performance Computing Center for letting us use their
computing resources.  This work was supported by NSF grant
Nos.~DMR-0908437, DMR-1307632.  CYD acknowledges support by the Singapore National
Research Foundation under grant No.~NRFF2012-02.
\end{acknowledgments}

\bibliography{references}

\end{document}